\documentclass[12pt]{article}
\usepackage{epsf,epsfig,amssymb,amsmath,graphicx,float}

\newcommand{\lsim}{\raisebox{-0.13cm}{~\shortstack{$<$ \\[-0.07cm] $\sim$}}~}
\newcommand{\gsim}{\raisebox{-0.13cm}{~\shortstack{$>$ \\[-0.07cm] $\sim$}}~}
\newcommand{\GeV}{\ensuremath{{\rm GeV}}}

\begin{document}
\renewcommand{\thefootnote}{\fnsymbol{footnote}}

\begin{titlepage}

\begin{center}

\vspace{1cm}

{\Large {\bf Relic Abundance of Asymmetric Dark Matter }}

\vspace{1cm}

{\bf Hoernisa Iminniyaz}$^a$\footnote{wrns@xju.edu.cn},
{\bf Manuel Drees}$^b$\footnote{drees@th.physik.uni-bonn.de} \ \ \ and \\
{\bf Xuelei Chen}$^c$\footnote{xuelei@cosmology.bao.ac.cn} \\
	
\vskip 0.15in
{\it
$^a${School of Physics Science and Technology, Xinjiang University, \\
Urumqi 830046, China} \\
$^b${Bethe Center for Theoretical Physics and Physikalisches Institut,
  Universit\"at Bonn, Nussallee 12, 53115 Bonn, Germany}\\
$^c${National Astronomical Observatories, Chinese Academy of
    Sciences,\\ Beijing 100012, China}\\
}

\abstract{We investigate the relic abundance of asymmetric Dark Matter
  particles that were in thermal equilibrium in the early
  universe. The standard analytic calculation of the symmetric Dark
  Matter is generalized to the asymmetric case. We calculate the
  asymmetry required to explain the observed Dark Matter relic
  abundance as a function of the annihilation cross section. We show
  that introducing an asymmetry always reduces the indirect detection
  signal from WIMP annihilation, although it has a larger annihilation
  cross section than symmetric Dark Matter. This opens new
  possibilities for the construction of realistic models of MeV Dark
  Matter.}

\end{center}
\end{titlepage}
\setcounter{footnote}{0}

\section{Introduction}

One of the most striking features of cosmology today is that the
Universe contains a large amount of Dark Matter, whose mass density
exceeds that of the known baryonic matter by about five times.  Cosmic
microwave background (CMB) anisotropy observations by the Wilkinson
Microwave Anisotropy Probe (WMAP) yield an accurate determination of
the total amount of the Dark Matter \cite{wmap},
\begin{eqnarray} \label{wmap}
  \Omega_{\rm DM} h^2 = 0.1109 \pm 0.0056\, ,
\end{eqnarray}
where $\Omega_{\rm DM}$ is the Dark Matter (DM) density in units of the
critical density, and $h = 0.710 \pm 0.025$ is the Hubble constant in
units of 100 km sec$^{-1}$ Mpc$^{-1}$.

Although we already have a relatively precise measurement of the Dark
Matter density, we still do not know what the Dark Matter is.
Neutral, long--lived or stable weakly interacting massive particles
(WIMPs) are currently considered as excellent candidates for Dark
Matter. One of the most popular WIMP Dark Matter candidate is the
neutralino in supersymmetric theory, which is stable due to the
conserved $R$--parity \cite{review}.  Neutralinos are Majorana
particles, i.e. they are their own anti--particles. However, this is
only one possibility and we do not actually have any evidence that the
Dark Matter consists of Majorana particles. Indeed, most of the known
elementary particles are not Majorana particles, but rather have
distinct anti--particles; moreover, the ordinary matter in the
Universe is almost entirely made from baryons, with anti--baryons
contributing only a tiny fraction of the baryons. It is thus natural
to consider asymmetric Dark Matter (ADM), for which particles and
anti--particles are not identical. This allows scenarios where the
universe exhibits a Dark Matter asymmetry, i.e. there are more Dark
Matter particles than anti--particles (or vice versa).

One motivation for considering this scenario is that the average
density of baryons is comparable to that of Dark Matter; with
$\Omega_b \approx 0.046$, they differ only by a factor of about
five. This led to speculations that a common mechanism might give rise
to the known baryon asymmetry and a postulated asymmetry in the Dark
Matter sector \cite{adm-models,frandsen}.

One attractive feature of the WIMP Dark Matter scenario is that it
provides a natural explanation for the observed Dark Matter abundance
(the ``WIMP miracle'').  At early time when the interaction rates are
high, the WIMP particles are in thermal equilibrium with the rest of
the cosmic fluid. As the Universe expands and the temperature of the
fluid drops, the interaction rates become smaller. Eventually the Dark
Matter particles decouple from the rest of the cosmic fluid, so that
their co--moving density is constant (``frozen''). For symmetric Dark
Matter, the resulting relic abundance is of the same order of
magnitude as the observed value, if the WIMP annihilation cross
section is of weak size. This hints at a connection between
weak--scale physics (which is also being probed at colliders) and Dark
Matter.

The abundance of Dark Matter can be calculated by solving the
Boltzmann equation which describes the time evolution of particle
densities in the expanding Universe. In case of symmetric Dark Matter,
such calculations have been done for various standard
\cite{standard-cos,dkk} and nonstandard \cite{nonstandard-cos,dik}
cosmological scenarios.

In this work we generalize this calculation to the case of asymmetric
Dark Matter. We solve the relevant Boltzmann equations both
numerically and in an analytic approximation. We assume that WIMPs can
annihilate only with their anti--particles, so that the number of
particles minus the number of anti--particles in a co--moving volume
is conserved. Without loss of generality we assume that there are more
particles than anti--particles.\footnote{In other words, we define the
  ``particle'' to be the one with the larger density, if an asymmetry
  exists.} We assume that the Dark Matter asymmetry, if any, is
created well before Dark Matter annihilation reactions freeze
out. Moreover, we assume standard cosmology during and after WIMP
decoupling. This implies a constant co--moving entropy density, and
absence of late non--thermal WIMP production (e.g. from the decay of
heavier particles).
 
This paper is arranged as follows. In Sec.~2, we discuss the Boltzmann
equations and the relic abundance of asymmetric Dark Matter. In
Sec.~3, we derive an approximate analytical formula for the asymmetric
case. In general the resulting Dark Matter density depends both on the
Dark Matter asymmetry and on the Dark Matter annihilation cross section.
In Sec.~4, we obtain constraints on these parameters from the observed
Dark Matter relic abundance. We also compute the ratio of anti--particle
and particle densities in the allowed part of parameter space, and
show that a non--vanishing asymmetry leads to a reduced rate of WIMP
annihilation in galactic haloes. Finally, in Sec.~5 we summarize our
results and draw some conclusions.

\section{Relic Abundance of Asymmetric Dark Matter}

Consider a Dark Matter particle denoted by $\chi$ that is {\em not}
self--conjugate, i.e. the anti--particle $\bar\chi \neq \chi$. The relic
densities of $\chi$ and $\bar\chi$ are determined by solving the
Boltzmann equations which describe the time evolution of the number
densities $n_{\chi}$, $n_{\bar\chi}$ in the expanding universe. Under
the assumptions that only $\chi \bar \chi$ pairs can annihilate into
Standard Model (SM) particles, while $\chi\chi$ and $\bar \chi \bar
\chi$ pairs cannot, the relevant Boltzmann equations are:
\begin{eqnarray} \label{eq:boltzmann_n}
\frac{{\rm d}n_{\chi}}{{\rm d}t} + 3 H n_{\chi} &=&  - \langle \sigma v\rangle
  (n_{\chi} n_{\bar\chi} - n_{\chi,{\rm eq}} n_{\bar\chi,{\rm eq}})~;
  \nonumber \\
\frac{{\rm d}n_{\bar\chi}}{{\rm d}t} + 3 H n_{\bar\chi} &=&  
   - \langle \sigma v\rangle (n_{\chi} n_{\bar\chi} - n_{\chi,{\rm
       eq}} n_{\bar\chi,{\rm eq}})~.
\end{eqnarray}
During the radiation dominated epoch, the expansion rate is given by 
\begin{equation} \label{H}
H = \frac{\pi T^2}{M_{\rm Pl}} \sqrt{\frac{g_*}{90}}\,,
\end{equation}
where $M_{\rm Pl} = 2.4\times 10^{18}$ GeV is the reduced Planck mass,
and $g_*$ is the effective number of the relativistic degrees of
freedom.  Here we will consider only the case of cold Dark Matter,
i.e. we assume the Dark Matter particles were already
non--relativistic at decoupling. The equilibrium number densities
$n_{\chi,{\rm eq}}$ and $n_{\bar\chi,{\rm eq}}$ are then given by
\begin{eqnarray} \label{n_eq}
  n_{\chi,{\rm eq}} &=& g_\chi ~{\left( \frac{m_\chi T}{2 \pi} \right)}^{3/2}
  {\rm e}^{(-m_\chi + \mu_\chi)/T}, \nonumber \\
  n_{\bar\chi,{\rm eq}} &=&  g_\chi ~{\left( \frac{m_\chi T}{2 \pi}
    \right)}^{3/2} {\rm e}^{(-m_\chi - \mu_\chi)/T},
\end{eqnarray}
where $m_\chi$ is the mass of the WIMP, $\mu_\chi$ is the chemical
potential of the particles, and $g_{\chi}$ is the number of the
internal degrees of freedom of the $\chi$ particle. We have used the
fact that $\mu_{\bar\chi} = -\mu_\chi$ in equilibrium. As a result,
the chemical potential drops out in the product $n_{\chi,{\rm eq}}
n_{\bar\chi,{\rm eq}}$. In Eqs.(\ref{eq:boltzmann_n}) the term
proportional to this product describes $\chi \bar \chi$ production
from SM particles; it is clear that this term should not be affected
by the WIMP chemical potential.\footnote{Here we ignore effects due to
  Bose enhancement or Fermi suppression in the {\em final} state, as
  usual in the treatment of the decoupling of WIMPs. These effects
  become important only if the Dark Matter particles were
  (semi--)relativistic at decoupling \cite{dkk}.}

We follow the standard picture of the Dark Matter particle
evolution. At high temperature the $\chi$ and $\bar\chi$ particles are
in thermal equilibrium in the early universe. When $T$ drops below the
mass $m_\chi$, the number densities $n_{\chi,{\rm eq}}$ and
$n_{\bar\chi,{\rm eq}}$ decrease exponentially, as long as $m_\chi >
|\mu_\chi|$. Eventually the interaction rates $\Gamma = n_{\bar\chi}
\langle \sigma v \rangle$ and $\bar{\Gamma} = n_\chi \langle \sigma v
\rangle$ therefore drop below $H$, which scales like $T^2$, see
Eq.(\ref{H}). The $\chi$ and $\bar\chi$ distributions are then no
longer kept in chemical equilibrium, and their co--moving number
densities approach constants. The temperature at which the WIMPs drop
out of chemical equilibrium is called the freeze--out temperature.

The Boltzmann equations (\ref{eq:boltzmann_n}) can be rewritten in
terms of the dimensionless quantities $Y_\chi = n_\chi/s$,
$Y_{\bar\chi} = n_{\bar\chi}/s$, and $x = m_\chi/T$, where
\begin{equation} \label{s}
s= (2 \pi^2/45) g_* T^3
\end{equation}
is the entropy density. If we assume that the universe expands
adiabatically during this period, the Boltzmann equations become
\begin{equation} \label{eq:boltzmann_Y}
\frac{d Y_{\chi}}{dx} = - \frac{\lambda \langle \sigma v \rangle}{x^2}~
(Y_{\chi}~ Y_{\bar\chi} - Y_{\chi, {\rm eq}}~Y_{\bar\chi, {\rm eq}}   )\,;
\end{equation}
\begin{equation} \label{eq:boltzmann_Ybar}
\frac{d Y_{\bar{\chi}}}{dx} 
= - \frac{\lambda \langle \sigma v \rangle}{x^2}~
 (Y_{\chi}~Y_{\bar\chi} - Y_{\chi, {\rm eq}}~Y_{\bar\chi, {\rm eq}} )\,,
\end{equation}
where we have introduced the constant
\begin{equation} \label{lambda}
\lambda = \frac{4 \pi }{\sqrt{90} }~m_{\chi} M_{\rm Pl}~\sqrt{g_*}\,.
\end{equation}

Subtracting Eq.(\ref{eq:boltzmann_Y}) from Eq.(\ref{eq:boltzmann_Ybar}),
we obtain
\begin{equation} \label{eq:YYbar}
\frac{d Y_{\chi}}{dx} - \frac{d Y_{\bar{\chi}}}{dx} = 0\,.
\end{equation}
This implies
\begin{equation}  \label{eq:c}
 Y_{\chi} - Y_{\bar\chi} = C\,,
\end{equation}
where $C$ is a constant, i.e. the difference of the co--moving densities
of the particles and anti--particles is conserved. This follows from our
assumption that $\chi$ and $\bar{\chi}$ only annihilate with each
other, which could e.g. be due to conservation of some (global)
charge.\footnote{$C$ is proportional to the initial (high$-T$) value
  of the $\chi-\bar\chi$ asymmetry, $A_\chi = (n_\chi - n_{\bar\chi})
  / (n_\chi + n_{\bar \chi})$. However, $A_\chi$ will change with time
  or temperature once the WIMPs become non--relativistic, approaching
  a (larger) constant again once the WIMPs decouple. We therefore
  prefer to use $C$ rather than $A_\chi$ to parameterize the $\chi -
  \bar\chi$ asymmetry.} Inserting Eq.(\ref{eq:c}) into
Eqs.(\ref{eq:boltzmann_Y}) and (\ref{eq:boltzmann_Ybar}), our
Boltzmann equations become 
\begin{equation} \label{eq:Yc}
\frac{d Y_{\chi}}{dx} = - \frac{\lambda \langle \sigma v \rangle}{x^2}~
      (Y_{\chi}^2 - C Y_{\chi} - P     )\, ;
\end{equation} 
\begin{equation} \label{eq:Ycbar}
\frac{d Y_{\bar{\chi}}}{dx} = - \frac{\lambda \langle \sigma v \rangle}{x^2}~
 (Y_{\bar\chi}^2 + C Y_{\bar\chi}  - P)\,, 
\end{equation}
where 
\begin{equation} \label{P}
P = Y_{\chi,{\rm eq}} Y_{\bar\chi,{\rm eq}} = (0.145
g_{\chi}/g_*)^2\,x^3\,{\rm e}^{-2x}\,. 
\end{equation} 
As noted above, $P$ does not depend on the chemical potential $\mu_\chi$,
which simplifies our calculation. These equations can be solved
numerically; a semi--analytical solution will be presented in the next
Section.

In most cases, the WIMP annihilation cross section can be expanded in
the relative velocity $v$ between the annihilating WIMPs\footnote{See
  ref.\cite{gs} for a discussion of scenarios where this expansion
  does not work.}
\begin{equation} \label{expand}
\sigma v = a + b v^2 + {\cal O}(v^4)\,.
\end{equation}
If $\chi \bar \chi$ annihilation from an $S-$wave initial state is
unsuppressed, the first term in Eq.(\ref{expand}) dominates,
while for annihilation from a $P-$wave initial state, $a=0$ but in
general $b\neq 0$. In all examples we know, annihilation from either
the $S-$ or the $P-$wave (or both) is allowed; the expansion
(\ref{expand}) is then sufficient for an accuracy of a few percent.

\begin{figure}[h!]
\begin{center}
\includegraphics[scale=0.9]{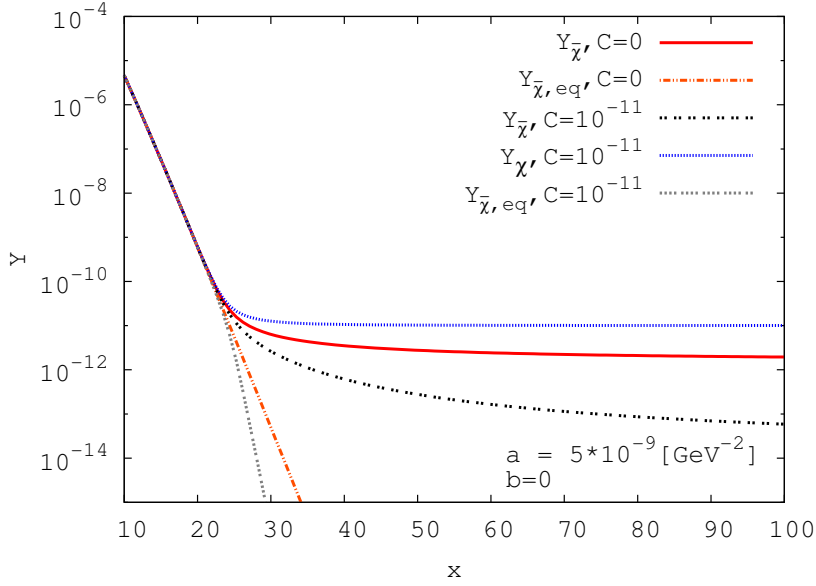}
\caption{\label{fig:b}{\footnotesize The evolution of the scaled
    $\chi$ and $\bar{\chi}$ abundances as function of $x=m/T$ for
$a=5\times 10^{-9}$ GeV$^{-2}$, $b=0, \ m = 100$ GeV and $C=10^{-11}$
or zero. The $\bar\chi$ equilibrium distributions are shown for
comparison. }}
\end{center}
\end{figure}

In Fig.~\ref{fig:b} we show the evolution of the relic abundances for
asymmetric Dark Matter $\chi$ and its anti--particle $\bar{\chi}$ as
function of $x=m/T$.\footnote{Similar numerical results have been
  presented in ref,~\cite{frandsen},} Here we take $b=0$ for
simplicity, with $a = 5\times 10^{-9} \GeV^{-2}$ and
$C=10^{-11}$. Results with non--vanishing $b$ are qualitatively the
same. For comparison, we also show result for the symmetric case
($C=0$). We show the actual as well as equilibrium values of the
$\bar\chi$ density; for $C = 0$ ($C > 0$), the equilibrium $\chi$
density is equal to (larger than) the equilibrium $\bar\chi$ density
shown in the figure. Note that $\chi$ and $\bar\chi$ are distinct
particles even if $C=0$.

As we can see in the figure, at high temperature (small $x$) $Y_\chi$
and $Y_{\bar\chi}$ are very close to their equilibrium values. This is
true irrespective of the initial value $Y_\chi(x_0)$ as long as $x_0$
is somewhat smaller than the freeze--out value $x_F$, $x_0 - x_F \gsim
1$ \cite{dik}. Moreover, $Y_{\chi,{\rm eq}}(x)$ is almost independent
of $C$ as long as $Y_{\chi,{\rm eq}}(x; \, C=0) > C$. This inequality
is satisfied for
\begin{equation} \label{C-cond}
x < x_C \simeq - \ln(C') + \frac{3}{2} \ln [- \ln(C')]\,,
\end{equation}
with $C' = 2^{5/2} \pi^{7/2} g_* C / (45 g_\chi) \simeq 310 C$; in the
last step we have taken $g_\chi=2, \, g_* = 90$. Numerically, $x_C
\simeq 24$ for $C = 10^{-11}$. $x_C$ is therefore also close to the
value where the $\bar \chi$ density begins to differ significantly
from its equilibrium value, which signals the on--set of
decoupling. For $C=0$ the actual $\bar\chi$ density then rather
quickly approaches a constant.

For $C = 10^{-11}$ the $\bar{\chi}$ abundance begins to deviate from
its equilibrium value also at $x \approx 23$. However, it keeps
decreasing for a large range of $x-$values, only slowly approaching a
smaller constant value ($\sim 10^{-13}$) at very large $x$. The
evolution of the $\chi$ abundance is even more remarkable: we have not
plotted its equilibrium density in this figure, because for $C =
10^{-11}$ it almost coincides with the actual $\chi$ density at all
temperatures! The ratio $Y_\chi / Y_{\chi,{\rm eq}}$ never exceeds
1.85; this maximum value is reached for $x \simeq 25$. This can be
understood from the observation that the $\chi$ density becomes nearly
equal to its equilibrium value, simply given by $C$, also at very
large $x$, where the $\bar\chi$ density is very small.

This large difference in the evolution of the $\chi$ and $\bar\chi$
densities originates again from our assumption that only $\chi \bar
\chi$ pairs can annihilate. Since $Y_\chi$ is bounded from below by
$C$, $\bar\chi$ particles still find a significant density of partners
for annihilation even after the nominal decoupling temperature. This
explains why $Y_{\bar\chi}$ keeps decreasing even at rather large
$x$. This in turn means that $\chi$ particles find even fewer partners
for annihilation than in the symmetric case ($C=0$). The $\chi$
density therefore ``decouples'' (i.e., $\Gamma_\chi < H$) even earlier
than for $C=0$, although its number density never differs very much
from its equilibrium value, as we just saw.

\begin{figure}[h!]
  \begin{center}
    \hspace*{-0.5cm} \includegraphics*[width=9cm]{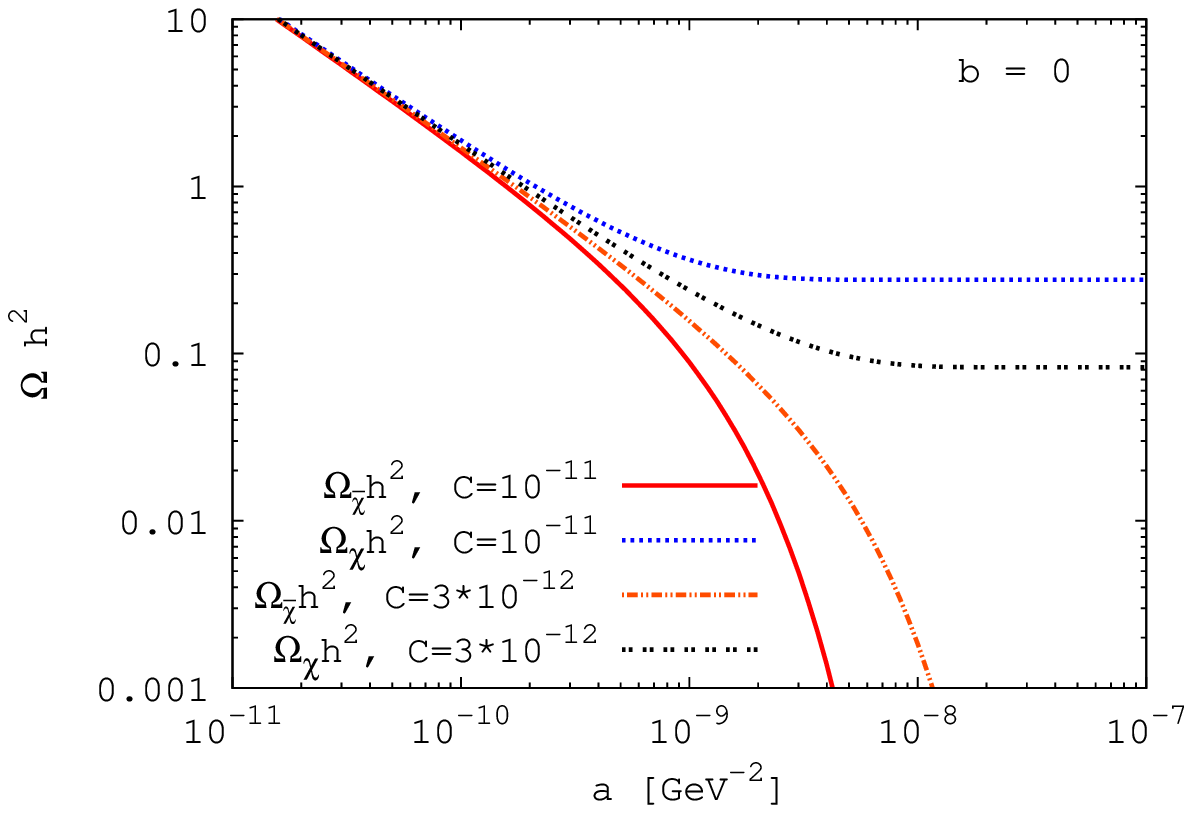}
    \put(-115,-12){(a)}
    \hspace*{-0.5cm} \includegraphics*[width=9cm]{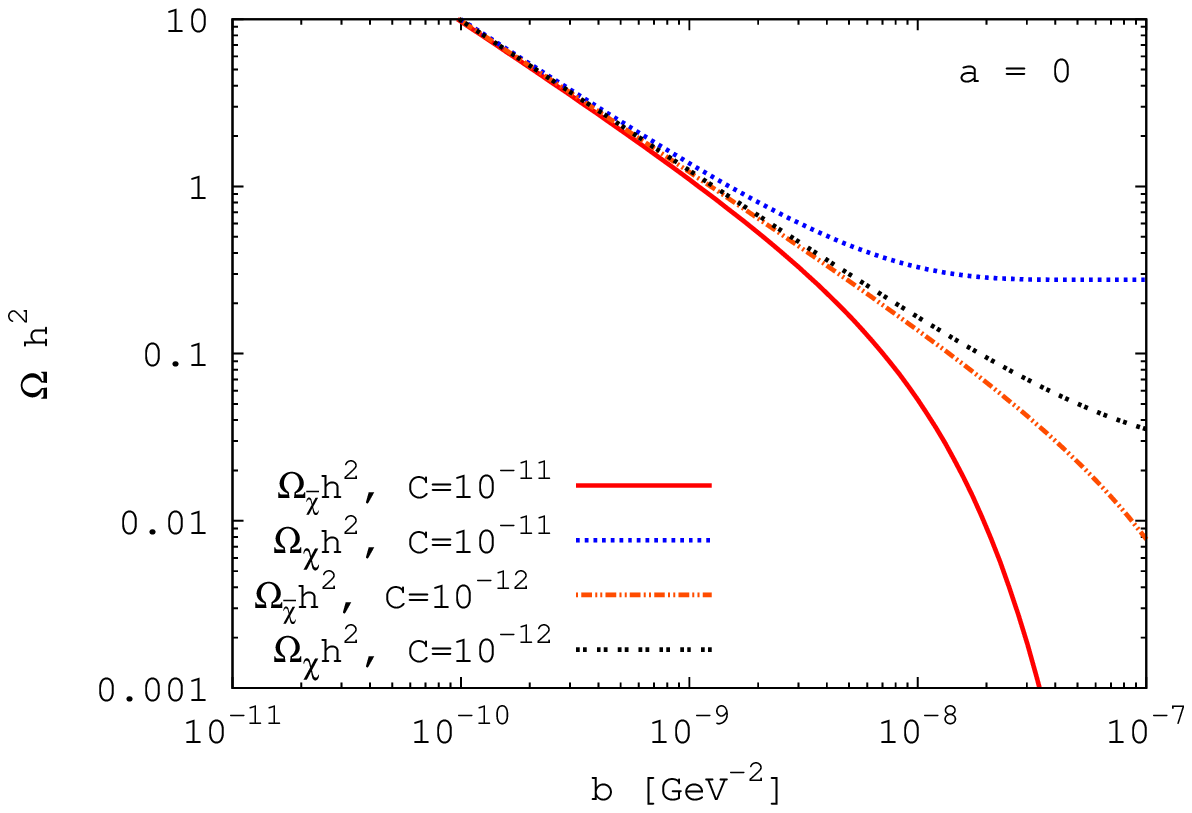}
    \put(-115,-12){(b)}
    \caption{\label{fig:a} \footnotesize The relic density $\Omega
      h^2$ for particle $\chi$ and anti--particle $\bar\chi$ as a
      function of the cross section. Here we take $m_\chi = 100$ GeV,
      $g_{\chi} = 2$ and $g_* = 90$. Panel (a) is for $b=0$, while
      panel (b) is for pure $P-$wave annihilation ($a=0$).}
  \end{center}
\end{figure}

With this understanding of the asymmetric Dark Matter thermal
decoupling, we now look at how its abundance depends on the
annihilation cross section. In Figure \ref{fig:a}(a,b) we plot the
relic density $\Omega h^2$ for particle $\chi$ and anti--particle
$\bar\chi$ as a function of the annihilation cross section, with two
different values of $C$. In panel (a) we assume a constant ($S-$wave)
cross section ($b=0$), while panel (b) is for a pure $P-$wave case
($a=0$).

We see that for small $\chi \bar\chi$ annihilation cross section the
relic density is independent of the asymmetry. This is the case
whenever the relic density in the symmetric case ($C=0$) is much
larger than the assumed value of $C$; in this situation the small
asymmetry clearly does not play much of a role in the decoupling
dynamics. Since for $C=0$ the relic density scales essentially
inversely with the annihilation cross section, this scenario
corresponds to small cross sections. More precisely, the product of
annihilation cross section and asymmetry determines whether the
asymmetry has significant impact on the final relic density or not.

In the regime of small cross section the velocity dependence of the
cross section is relevant. In the symmetric case the contribution of
the $b-$term in Eq.(\ref{expand}) to the final relic density is
suppressed by a factor $x_F/3 \simeq 7$ relative to that of the
$a-$term. A similar scaling holds in Fig.~\ref{fig:a} for small cross
section. 

On the other hand, as already shown in Fig.~\ref{fig:b}, for
sufficiently large annihilation cross section a non--vanishing
asymmetry $C$ does greatly affect the relic density. The scaled $\chi$
density $Y_\chi$ will then simply approach $C$, while the final $\bar
\chi$ density becomes insignificant. This is the limit of extremely
asymmetric dark matter. The total DM relic density then becomes
independent of the value and the velocity dependence of the cross
section (although the velocity dependence does play a role in deciding
what cross sections are ``large'' in this sense).

\section{Analytical Solution}

While the Boltzmann equations (\ref{eq:Yc}), (\ref{eq:Ycbar}) can be
solved numerically, it is still useful to have some analytical
solution of these equations, as in the symmetric Dark Matter case. In
our semi--analytical treatment we closely follow the standard
treatment \cite{standard-cos} of symmetric Dark Matter to solve
Eq.(\ref{eq:Ycbar}) for the $\bar \chi$ density. The $\chi$ density
can then be computed trivially using Eq.(\ref{eq:c}).

We begin by introducing the quantity $\Delta_{\bar\chi} = Y_{\bar\chi} -
Y_{\bar\chi,{\rm eq}}$. Its evolution follows directly from the
Boltzmann equation (\ref{eq:Ycbar}):
\begin{equation} \label{eq:delta}
\frac{d \Delta_{\bar\chi}}{dx} = - \frac{d Y_{\bar\chi,{\rm eq}}}{dx} - 
\frac{\lambda \langle \sigma v \rangle}{x^2}~
\left[\Delta_{\bar\chi}(\Delta_{\bar\chi} + 2 Y_{\bar\chi,{\rm eq}})  
      + C \Delta_{\bar\chi}   \right]\, .
\end{equation}

We consider the behavior of the solution of this equation in two
regimes. At sufficiently high temperature $Y_{\bar\chi}$ tracks its equilibrium
value $Y_{\bar\chi,{\rm eq}}$ very closely. In that regime $\Delta_{\bar\chi}$ 
is small, and $d \Delta_{\bar\chi}/dx$ and $\Delta_{\bar\chi}^2$ are
negligible. The Boltzmann equation (\ref{eq:delta}) then becomes
\begin{equation}\label{eq:delta_simp}
     \frac{d Y_{\bar\chi,{\rm eq}}}{dx}   =  - 
      \frac{\lambda \langle \sigma v \rangle}{x^2}~
      \left(2 \Delta_{\bar\chi} Y_{\bar\chi,{\rm eq}} + 
       C \Delta_{\bar\chi} \right)\,.
\end{equation}
In order to calculate the left--hand side of Eq.(\ref{eq:delta_simp}),
we need an explicit expression for the equilibrium density
$Y_{\bar\chi,{\rm eq}}(x)$. By definition the right--hand sides of the
Boltzmann equations (\ref{eq:boltzmann_Y}) and
(\ref{eq:boltzmann_Ybar}) vanish in equilibrium. Hence the right--hand
side of Eq.(\ref{eq:Ycbar}) should vanish as well for $Y_{\bar\chi} =
Y_{\bar\chi,{\rm eq}}$, which implies
\begin{equation} \label{eq:equili}
Y_{\bar\chi,{\rm eq}}^2 + C Y_{\bar\chi,{\rm eq}} - P   = 0\,.
\end{equation} 
This quadratic equation has two solutions, but only one of them yields
a positive $\bar\chi$ equilibrium density:
\begin{equation} \label{eq:bar_eq}
      Y_{\bar\chi,{\rm eq}} = - \frac{C}{2} + \sqrt{\frac{C^2}{4} + P}\,.
\end{equation}
Plugging Eq.(\ref{eq:bar_eq}) into Eq.(\ref{eq:delta_simp}) and
ignoring $x$ relative to $x^2$, we have 
\begin{equation} \label{bardelta_solu}
      \Delta_{\bar\chi} \simeq \frac{2 x^2 P}
      {\lambda \langle \sigma v \rangle\,(C^2 + 4 P)}\,.  
 \end{equation}
This solution will be used to determine the freeze--out temperature for 
$\bar{\chi}$, which we denote by $\bar{x}_F$. 

At sufficiently low temperature, i.e. for $x > \bar x_F$, we can
ignore the production term $\propto Y_{\bar\chi,{\rm eq}}$ in the
Boltzmann equation (\ref{eq:delta}), so that
\begin{equation} \label{eq:delta_late}
\frac{d \Delta_{\bar\chi}}{dx} = - \frac{\lambda \langle \sigma v  
\rangle}{x^2} \left( \Delta_{\bar\chi}^2 + C \Delta_{\bar\chi}    \right)\,.
\end{equation}
At this point we may assume $\Delta_{\chi}(\bar{x}_F) \gg
\Delta_{\chi}(\infty) $. Integrating Eq.(\ref{eq:delta_late}) from
$\bar{x}_F$ to $\infty$ then yields
\begin{equation} \label{eq:barY_solu}
      Y_{\bar\chi}(x \rightarrow \infty) = \frac{C}
       {\exp\left(C \lambda  \int^{\infty}_{\bar{x}_F} 
      \langle \sigma v \rangle\, x^{-2} dx\right) -1}.
\end{equation}
If we use the non--relativistic expansion (\ref{expand}) of the cross
section times relative velocity, the thermal averaging gives
\begin{equation} \label{eq:cross}
   \langle \sigma v \rangle = a + 6\,b x^{-1} + {\cal O}(x^{-2})\, .
\end{equation}
The solution (\ref{eq:barY_solu}) then becomes 
\begin{equation} \label{eq:barY_cross}
Y_{\bar\chi}(x \rightarrow \infty) =  \frac{C}
 { \exp \left[ C (4 \pi/ \sqrt{90})\, m_{\chi} M_{\rm Pl}\, 
 \sqrt{g_*} \, (a~ \bar{x}_F^{-1} + 3 b~ \bar{x}_F^{-2})   \right] -1}\,,
\end{equation}
where we have used Eq.(\ref{lambda}). Note that for small asymmetry
$C$, or more precisely for small argument of the exponential in the
denominator, $C$ will drop out, and the $\bar\chi$ relic density will
be inversely proportional to the annihilation cross section. This
reproduces the standard result \cite{standard-cos}. On the other hand,
Eqs.(\ref{eq:barY_solu}) or (\ref{eq:barY_cross}) indicate that the
$\bar\chi$ relic density will become exponentially small if the
product of annihilation cross section and asymmetry becomes large.

We can repeat this exercise to calculate the relic density of $\chi$
particles. The result is
\begin{equation} \label{eq:Y_cross}
       Y_{\chi}(x \rightarrow \infty) =   \frac{C}
       { 1 - \exp \left[- C (4 \pi/ \sqrt{90})\, m_{\chi} M_{\rm Pl}\, 
        \sqrt{g_*}  \, (a x_F^{-1} + 3 b x_F^{-2})  \right]}\,,
\end{equation}
where $x_F$ is the inverse scaled freeze--out temperature of
$\chi$. Note that Eqs.(\ref{eq:barY_cross}) and (\ref{eq:Y_cross}) are
only consistent with the constraint (\ref{eq:c}) if $x_F = \bar
x_F$. Recall from the discussion in the previous Section that for
sizable $C$ the reaction rate $\Gamma_\chi$ is (much) smaller than the
rate $\Gamma_{\bar \chi}$, since $\chi$ particles find (far) fewer
partners for annihilation. The fact that our treatment nevertheless
requires equal ``freeze--out'' temperatures for $\chi$ and $\bar \chi$
particles indicates that the intuitive condition $\Gamma(x_F) \simeq
H(x_F)$, i.e. reaction rate $\simeq$ expansion rate, can no longer be
applied consistently for the decoupling of asymmetric Dark Matter.

For convenience, we express the final abundance in terms of
\begin{equation}
\Omega_\chi h^2 =\frac{m_\chi s_0 Y_{\chi}(x \to \infty) h^2}{\rho_{\rm
  crit}}\,,
\end{equation}
where $s_0 = 2.9 \times 10^3~{\rm cm}^{-3}$ is the present
entropy density, and $\rho_{\rm crit} = 3 M_{\rm Pl}^2 H_0^2$ is the present
critical density.
The corresponding prediction for the present relic density for Dark Matter is 
given by
\begin{equation} \label{omega}
\Omega_{\rm DM} h^2 = 2.76 \times 10^8~ m_\chi \left[ Y_{\chi}~(x
  \rightarrow \infty) + Y_{\bar\chi}~(x \rightarrow \infty) \right]
\mbox{GeV}^{-1}\, . 
\end{equation}
For small product of annihilation cross section and asymmetry the WIMP
mass $m_\chi$ drops out in the final result (\ref{omega}). In
contrast, if this product becomes large, the $\bar \chi$ relic density
is negligible while the scaled $\chi$ relic number density is simply
given by $C$; in this limit the final Dark Matter mass density is thus
proportional to the mass of the Dark Matter particle. In that regard
the Dark Matter mass density of very asymmetric WIMPs behaves like
that of hot Dark Matter, although our WIMPs remain cold Dark Matter.

The remaining task is to fix the freeze--out temperature. Here the
standard method \cite{standard-cos} needs some modification. We start
from the standard definition of $\bar x_F$ which assumes that
freeze--out occurs when the deviation $\Delta_{\bar\chi}$ is of the
same order as the equilibrium value of $Y_{\bar\chi}$:
\begin{equation} \label{xf1}
\xi Y_{\bar\chi,{\rm eq}}( \bar{x}_{F_0}) = \Delta_{\bar\chi}( \bar{x}_{F_0})\,,
\end{equation}
where $\xi$ is a numerical constant of order unity.  We adopt the
usual value \cite{standard-cos} $\xi = \sqrt{2} -1$. By comparing with
numerical solutions, we find that the standard treatment
under--predicts the $\bar\chi$ relic density for sizable asymmetry. To
very good approximation the ratio of the exact analytical solution and
the approximation using $\bar x_{F_0}$ as decoupling temperature only
depends on the product $\lambda C \langle \sigma v
\rangle(x_{F_0})/x^2_{F_0}$, rather than on $C, \, m_\chi$ and the
annihilation cross section separately. This is not so surprising,
since this expression appears as integrand in
Eq.(\ref{eq:barY_solu}). Moreover, the ratio of exact and approximate
solution remains of order unity as long as the above product is $\lsim
1$.

We therefore need only a relatively minor modification of the standard
treatment. After some trials, we found that the most elegant and
simple way to improve the analytical approximation is through a small
shift of the freeze--out temperature:
\begin{equation} \label{eq:xF}
\bar{x}_F = \bar{x}_{F_0}\, 
   \left(1 + \frac{0.285 \lambda\, a C}{\bar{x}^3_{F_0}}
+ \frac{1.350 \lambda\, b C}{\bar{x}^4_{F_0}} \right)\,.
\end{equation}

\begin{figure}[h!]
  \begin{center}
    \hspace*{-0.5cm} \includegraphics*[width=8.6cm]{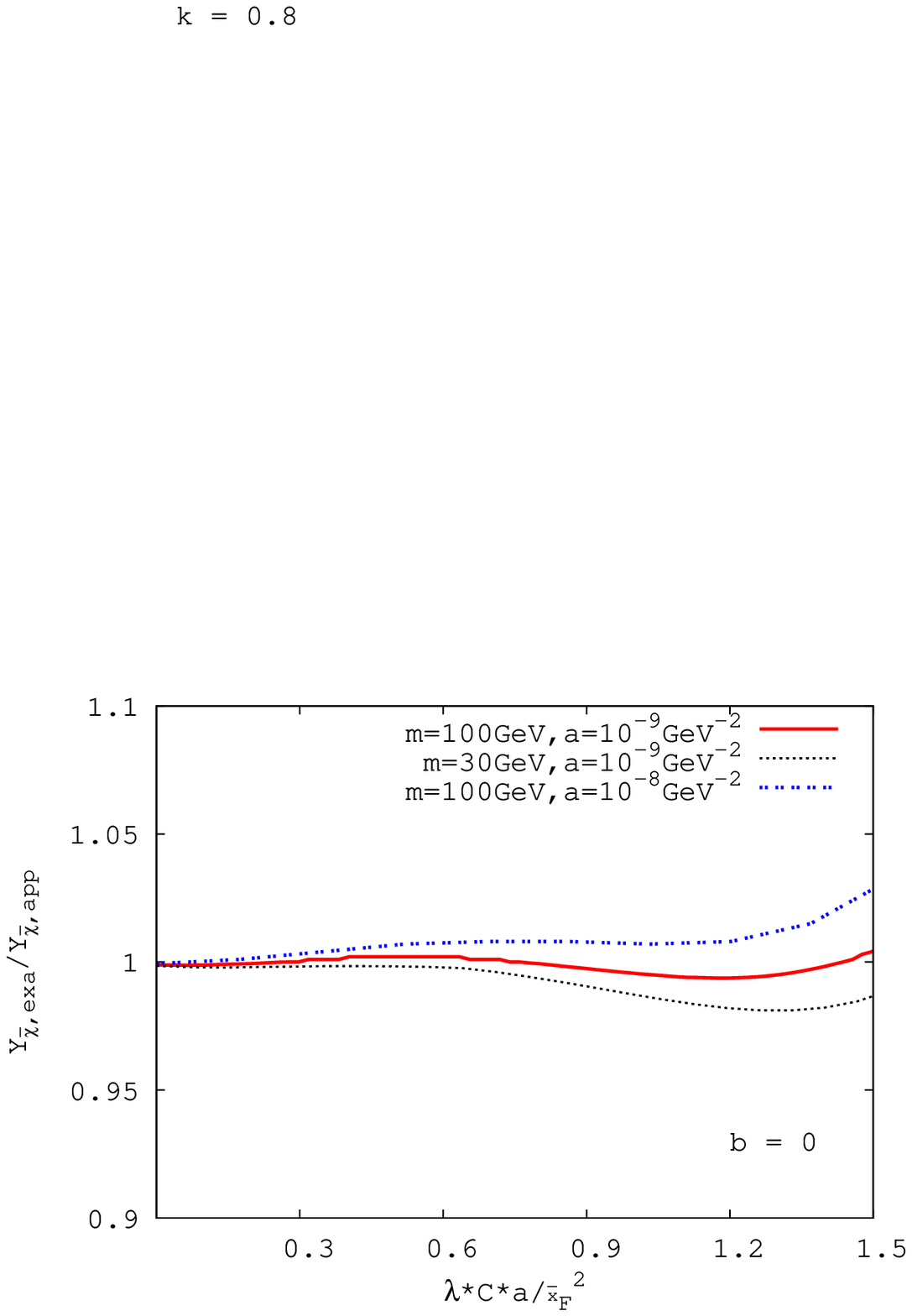}
    \hspace*{-0.5cm} \includegraphics*[width=8.6cm]{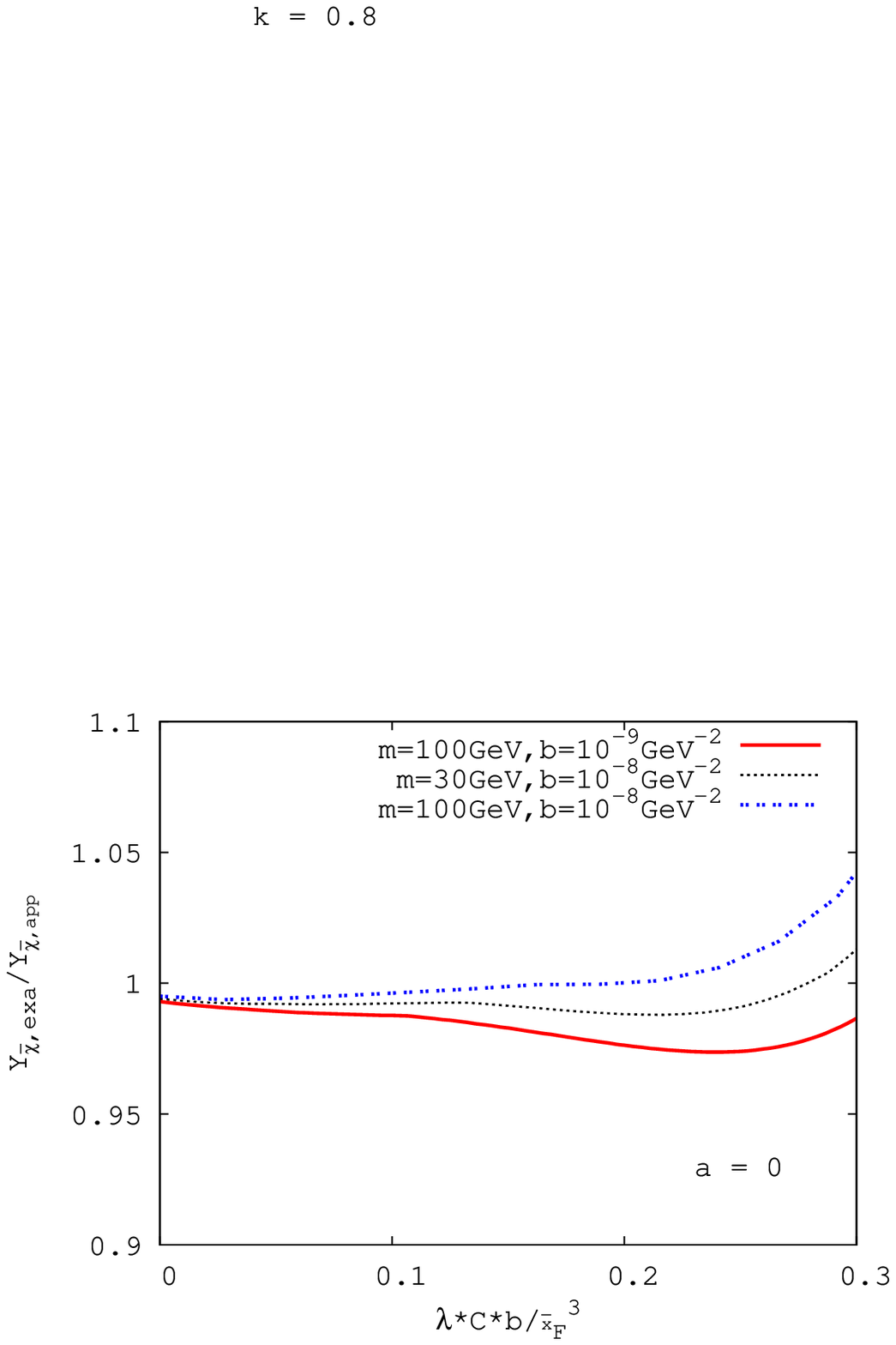}
    \caption{\label{fig:c} \footnotesize
    The ratio of the exact value of the $\bar\chi$
    particle abundance to the analytic value of $\bar\chi$ 
    particle abundance.  }
  \end{center}
\end{figure}

In Fig. \ref{fig:c} we plot the ratio of the exact value of the
$\bar\chi$ particle abundance to our analytical approximation. As
shown in the figure, the approximate analytic result matches the exact
numerical result very well as long as $\lambda C \langle \sigma v
\rangle(x_F)/x_F^2 \lsim 1$. For even larger product of cross section
and asymmetry the deviation becomes sizable again. However, in that
case the $\bar \chi$ relic density is entirely negligible.
Quantitatively, our approximation reproduces the exact numerical
solution to better than 2\% (5\%) as long as the final $\bar\chi$
relic density exceeds $10^{-4}$ ($10^{-9}$) times the final $\chi$
relic density.

Since for large product $\lambda C \langle \sigma v \rangle$ the
scaled $\chi$ relic number density is simply given by $C$, our simple
approximation reproduces the exact $\chi$ relic density, and also the
exact total WIMP ($\chi + \bar \chi$) mass density, to better than 2\%
for all combinations of parameters.

We finally note that the shift (\ref{eq:xF}) of $\bar{x}_F$ amounts to
$\lsim 2\%$ for the parameter range covered in Fig.~\ref{fig:c}. This
nevertheless can lead to a significant change of the $\bar\chi$
relic density due to the exponential dependence of
Eq.(\ref{eq:barY_cross}) on $\bar{x}_F$.


\section{Constraints on Parameter Space and Indirect Detection
  Signals} 
\setcounter{footnote}{0}

The Dark Matter abundance can be derived from cosmological
observations.  This measurement does not depend on the nature of dark
matter, e.g. whether it is symmetric or asymmetric, as long as the
Dark Matter particles are sufficiently slow (``cold'' Dark Matter).
Using only the CMB data, the WMAP team derived the Dark Matter density
given in Eq.(\ref{wmap}) for the minimal $\Lambda$CDM model. The
precision could be further improved by including observations of
baryon acoustic oscillation (BAO) and/or direct measurements of the
Hubble constant ($H_0$) into the fits \cite{wmap}. We note that
besides the statistical error quoted in Eq.(\ref{wmap}), there could
also be systematic errors, and the result depends somewhat on the
model and priors adopted. Nevertheless, we may use the measured Dark
Matter abundance to place constraints on the asymmetric Dark Matter
model. In view of the above remarks, we adopt the conservative range
\begin{equation} \label{range}
0.10 < \Omega_{\rm DM} h^2 < 0.12 
\end{equation}
when setting bounds on the model parameters; this is roughly the $\pm
2\sigma$ limit for the abundance indicated by the WMAP fit given in
Eq.(\ref{wmap}). Note that for asymmetric Dark Matter, the $\chi$ and
$\bar\chi$ contributions have to be added:
\begin{equation} \label{add}
\Omega_{\rm DM}=\Omega_\chi + \Omega_{\bar{\chi}}\,.
\end{equation}

In this Section we assume that the annihilation cross section is given
by Eq.(\ref{expand}). We use exact numerical solutions of the
Boltzmann equations; however, the analytical solution described in the
previous Section reproduces the exact result to better than 5\% for
all combinations of parameters we analyze in this Section.

\begin{figure}[h]
  \begin{center}
    \hspace*{-0.5cm} \includegraphics*[width=8.7cm]{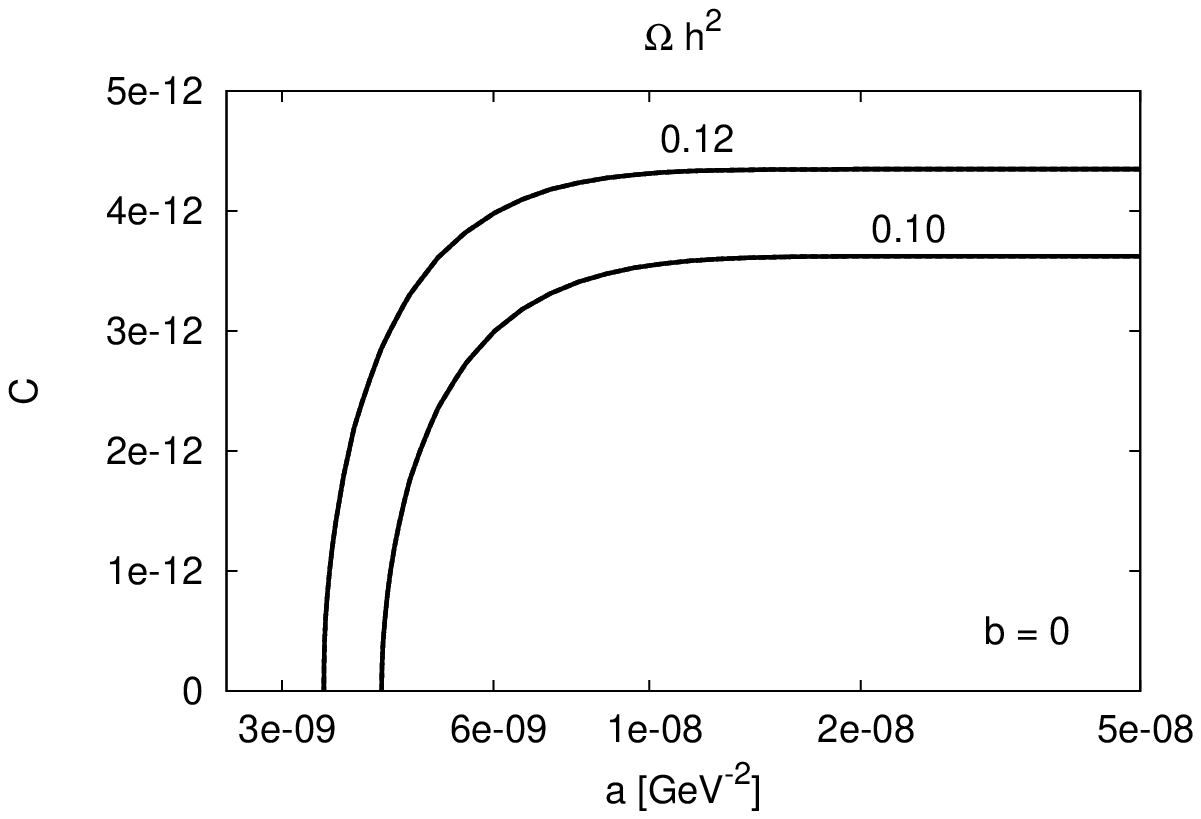}
    \hspace*{-0.5cm} \includegraphics*[width=8.7cm]{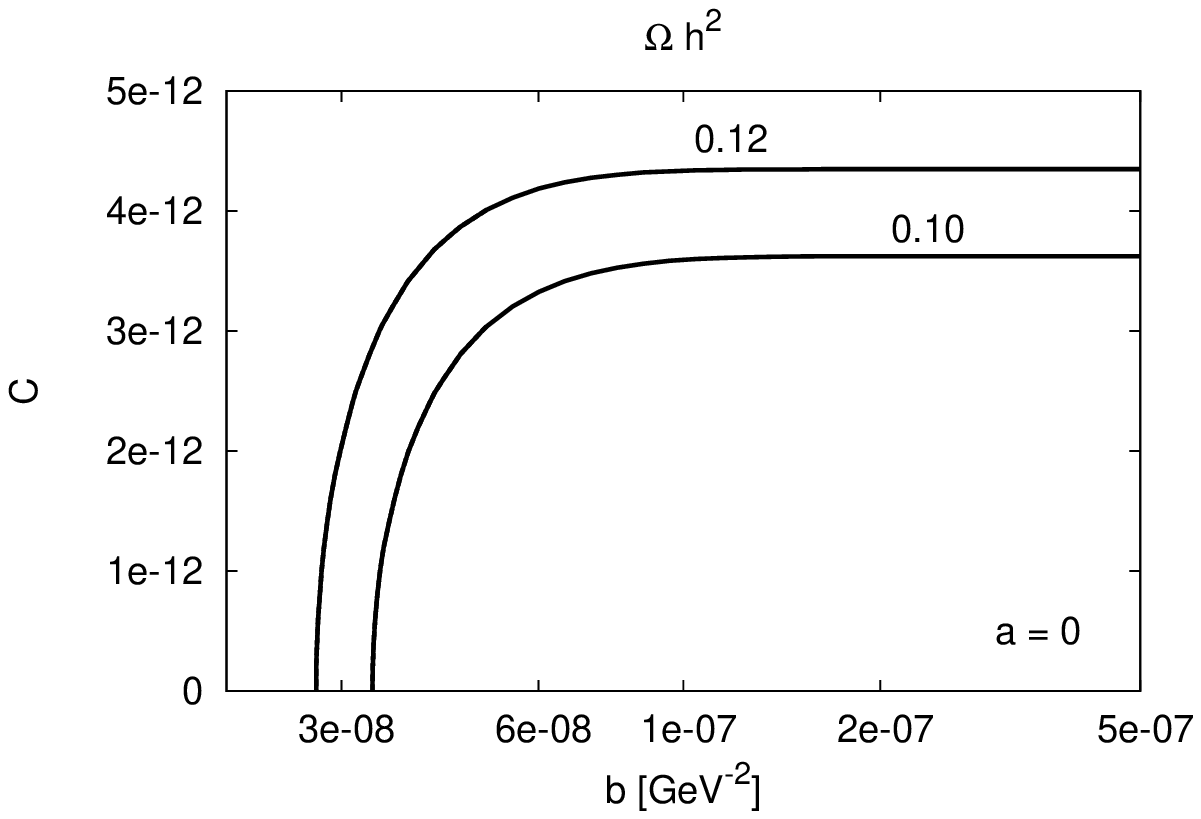}
    \caption{\label{fig:e} \footnotesize
    The allowed region in the $(a,C)$ plane for $b=0$ (left), and in
    the $(b,C)$ plane for $a=0$ (right), when the Dark Matter density
    $\Omega h^2$ lies between 0.10 and 0.12. Here we take $m_\chi = 100$ GeV, 
    $g_{\chi} = 2$ and $g_* = 90$; the allowed values of $C$ scale to
    good approximation like $1/m_\chi$.}
    \end{center}
\end{figure}

The relation between the cross section parameters $a$ and $b$ (for
$b=0$ and $a=0$, respectively) and the asymmetry $C$ are shown in
Fig.~\ref{fig:e} for two values of the total Dark Matter density. For
$C=0$, i.e. in the symmetric case, one needs $a \sim 4-5 \times
10^{-9}\ \GeV^{-2}$ for $b=0$ (i.e. $S-$wave dominance), and $b \sim 3
\times 10^{-8}\ \GeV^{-2}$ for $a=0$ ($P-$wave dominance). These cross
sections are about two times larger than the cross sections required
for self--conjugate (Majorana) Dark Matter with $g_\chi=2$ degrees of
freedom. This is due to the fact that the (equal) $\chi$ and
$\bar\chi$ contributions have to be added, as in Eq.(\ref{add}); each
of them is about as large as the contribution from Majorana dark
matter with the same cross section. The resulting doubling of the
total Dark Matter density has to be corrected by also (approximately)
doubling the annihilation cross section.\footnote{One can also compare
  non self--conjugate Dark Matter with Majorana Dark Matter with
  $g_\chi=4$. Note that for fixed annihilation cross section, the
  final relic density depends on $g_\chi$ only logarithmically (via
  $x_f$). An asymmetric Dark Matter scenario with given cross section
  then corresponds to a model of Majorana Dark Matter with half the
  cross section, since in the non self--conjugate case effectively
  only half the initial states contribute to the annihilation ($\chi
  \bar \chi$ and $\bar \chi \chi$ contribute but $\chi\chi$ and
  $\bar\chi \bar\chi$ do not).}

Now consider the asymmetric case, i.e. $C \neq 0$. For small values of
$C$, the abundance is not much affected, so the iso--abundance
contours initially rise almost vertically. This can be understood from
the analytical expressions (\ref{eq:barY_cross}) and
(\ref{eq:Y_cross}): if we ignore the shift of $\bar x_F$ given by
(\ref{eq:xF}), which is very small for small $C$, we find $Y_\chi(x
\rightarrow \infty) \simeq Y_\chi(x \rightarrow \infty;\, C=0) + C/2 +
{\cal O}(C^2)$ and $Y_{\bar \chi}(x \rightarrow \infty) \simeq
Y_\chi(x \rightarrow \infty;\, C=0) - C/2 + {\cal O}(C^2)$, i.e. the
correction to the total Dark Matter relic density begins at ${\cal
  O}(C^2)$.

For somewhat larger cross sections the curves quickly flatten out. A
larger cross section leads to smaller decoupling temperature, and
hence to a smaller relic density; this has to be compensated by
increasing $C$. Specifically, once the cross section is twice as
large as that required for vanishing asymmetry, the required asymmetry
is already only about 5\% smaller than the asymptotic value required
for large annihilation cross section (i.e., for strongly asymmetric
Dark Matter, where the $\bar\chi$ relic density is negligible and the
$\chi$ density is simply given by $C$). Since $C=0$ gives the minimal
allowed value of $\langle \sigma v \rangle$ while $\langle \sigma v
\rangle \longrightarrow \infty$ defines the maximal allowed value of
$C$, the above statement can be written as
\begin{equation} \label{cmax}
C(\langle \sigma v \rangle > 2 \langle \sigma v \rangle|_{\rm min}) >
0.95 C_{\rm max}\,.
\end{equation}

\begin{figure}[h!]
  \begin{center}
    \hspace*{-0.5cm} \rotatebox{270}{\includegraphics[width=6cm]{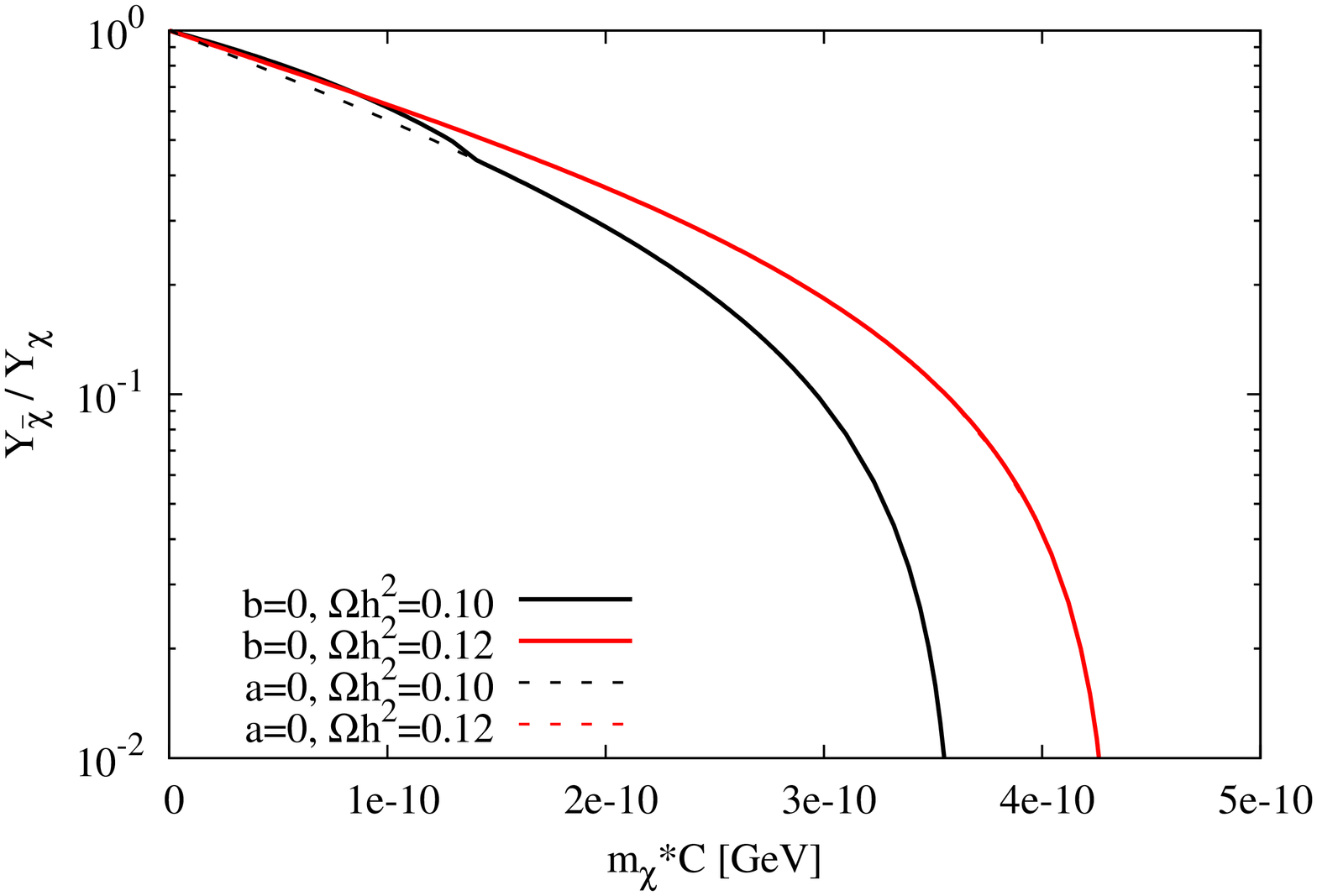}}
    \hspace*{-0.5cm} \rotatebox{270}{\includegraphics[width=6cm]{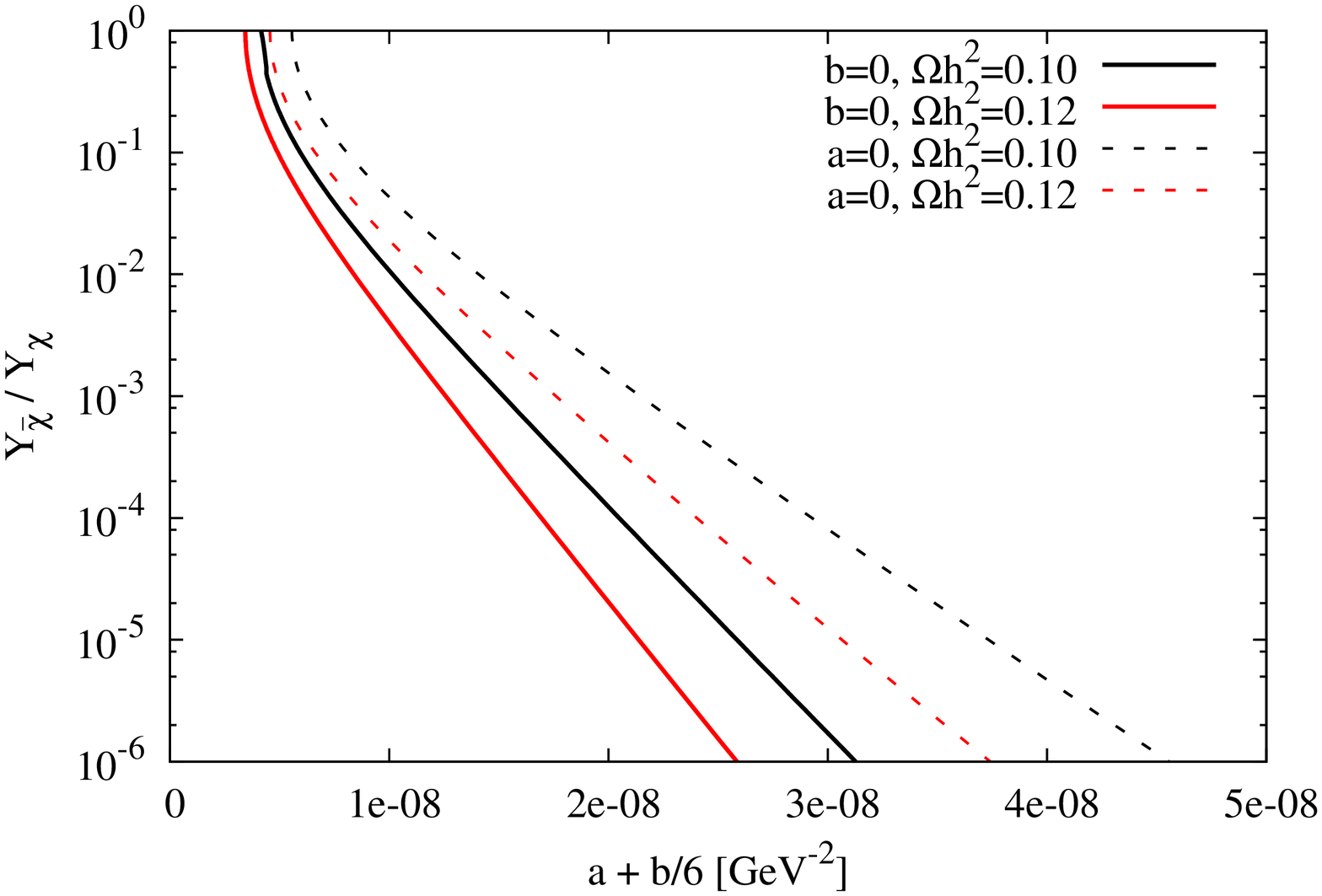}}
    \caption{\label{fig:f} \footnotesize Ratio of the current
      anti--particle abundance $Y_{\bar\chi}$ to the particle abundance
      $Y_\chi$ as a function of $m_\chi \cdot C$ (left) or the sum $a +
      b/6$ characterizing the annihilation cross section (right) for
      different combinations of cross sections and total Dark Matter
      relic density. In the left (right) frame the annihilation cross
      section (asymmetry $C$) is chosen such that the total dark
      matter density $\Omega_{\rm DM} h^2$ has the indicated
      value. Results are for $g_*=90$ and $m_\chi = 100$ GeV, but are
      almost independent of $m_\chi$.}
    \end{center}
\end{figure}

\begin{figure}[h!]
 \begin{center}
    \hspace*{-0.5cm} \rotatebox{270}{\includegraphics*[width=6cm]{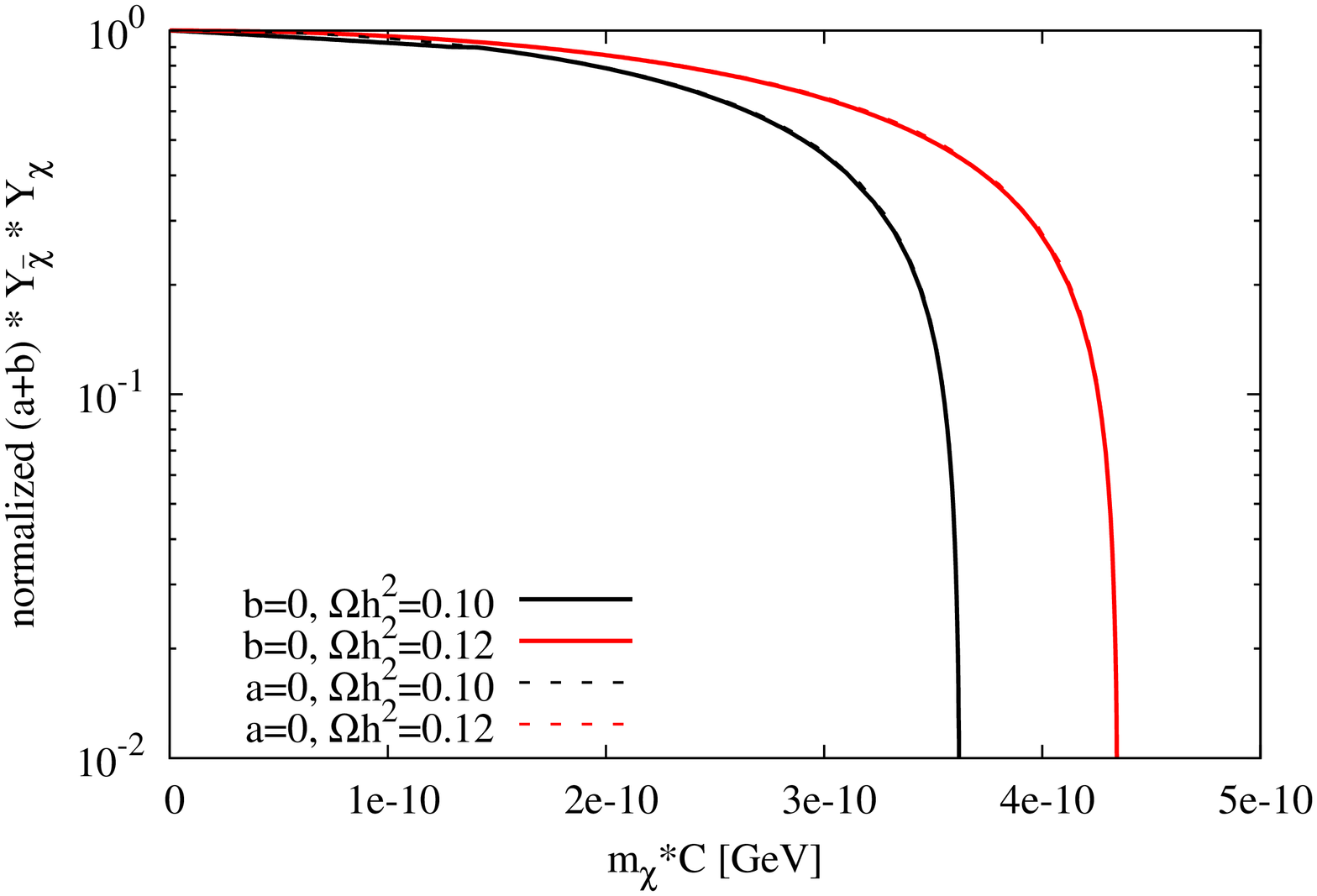}}
    \hspace*{-0.5cm} \rotatebox{270}{\includegraphics*[width=6cm]{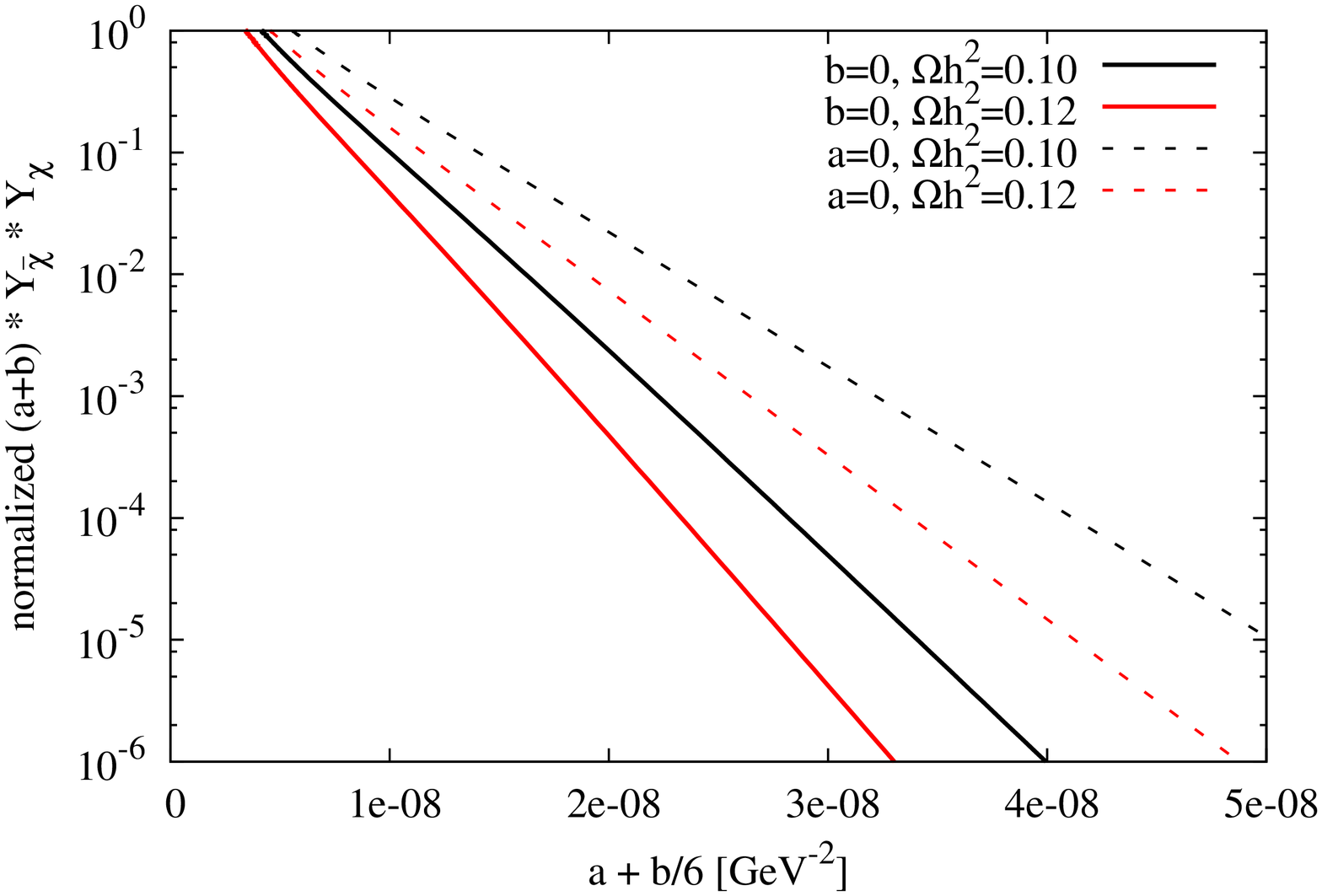}}
\caption{\label{fig:d} \footnotesize
Today's annihilation rate for asymmetric Dark Matter divided by the
same quantity for symmetric Dark Matter ($C=0$), as a function of
$m_\chi \cdot C$ (left) and $a + b/6$ (right). Parameters and
conventions are as in Fig.~\ref{fig:f}.} 
    \end{center}
\end{figure}

Of greater physical interest than the ratio $C/C_{\rm max}$ are the
fraction $Y_{\bar\chi} / Y_\chi$ as well as the product $\sigma Y_\chi
Y_{\bar\chi}$, where the particle densities are to be taken at the
present epoch, i.e. for very large $x$. These quantities are shown in
Figs.~\ref{fig:f} and \ref{fig:d}, respectively.

The ratio $Y_{\bar\chi} / Y_\chi$ determines the contribution of
anti--particles to the current Dark Matter density. In the left frame
of Fig.~\ref{fig:f} this ratio is shown as function of $m_\chi \cdot
C$, so that the results are nearly independent of $m_\chi$; the
annihilation cross section has been chosen such that the relic density
has the indicated value ($\Omega_{\rm DM} h^2 = 0.10 \ [0.12]$ for the
black [grey or red] curves). Plotted in this way, the result is almost
independent of whether the $a-$ or $b-$term in the annihilation cross
section (\ref{expand}) dominates.

We see that for small $C$ the logarithm of the ratio falls essentially
linearly with increasing $C$, with relatively gentle slope. Recall
from the discussion of Fig.~\ref{fig:e} that for constant annihilation
cross section the ratio would behave approximately like $[Y_\chi(C=0)
-C/2] / [Y_\chi(C=0)+C/2]$ for sufficiently small $C$; this gives
$\ln(Y_{\bar\chi} / Y_\chi) \simeq - C/Y_\chi(C=0)$.  Fig.~\ref{fig:e}
also showed that increasing $C$ requires at first only very small
increase of the annihilation cross section.

This behavior continues until $C$ reaches about 95\% of its maximal
value, where the ratio $Y_{\bar \chi} / Y_\chi \simeq 0.03$. When $C$
is increased even more the ratio plummets. This is caused by the
required increase of the annihilation cross section, which has to
become very large as $C \longrightarrow C_{\rm max}$. We saw in
Fig.~\ref{fig:a} that this strongly suppresses the $\bar\chi$ relic
density. The logarithmic slope of the ratio therefore approaches
$-\infty$ as $C \rightarrow C_{\rm max}$.

In this region of large asymmetry it therefore seems more useful to
consider the ratio as function of the annihilation cross section. This
is shown in the right frame of Fig.~\ref{fig:f}. We have chosen $a +
b/6$ as $x-$axis, so that the cases $a=0$ (dashed curves) and $b=0$
(solid) can be shown on the same scale. The asymmetry $C$ is chosen
such that the total relic density has the indicated value. We see that
the ratio first decreases very quickly when the annihilation cross
section is increased beyond its minimal possible value, which is
reached for $C=0$. This corresponds to the rapid increase of the
required asymmetry $C$ with increasing cross section shown in
Fig.~\ref{fig:e}. For cross sections larger than roughly twice the
minimal value the curves flatten out, and have essentially constant
logarithmic slope from then on; this slope depends on the desired
relic density and on the functional form of the cross section ($a-$ or
$b-$dominance).

The product $\sigma Y_\chi Y_{\bar\chi}$ determines the strength of
indirect Dark Matter detection signals from the annihilation of WIMPs
in the haloes of galaxies. In Fig.~\ref{fig:d} this product is
normalized to the same product taken at $C=0$, i.e. for symmetric Dark
Matter. Notice that the cross section in the normalization factor is
kept fixed, to the value required to get the indicated Dark Matter
relic density, while the cross section in the numerator is allowed to
vary. Gravity acts the same way on particles and anti--particles, and
the WIMP distribution in the present universe is basically only
determined by gravity.\footnote{Electromagnetic interactions may
  affect the distribution of baryonic matter, which in turn can affect
  the Dark Matter distribution through gravitational interactions.}
The ratio $n_{\bar \chi} / n_\chi$ should therefore be equal to the
universally averaged $Y_{\bar\chi} / Y_\chi$ everywhere.  This
normalized product thus shows how the strength of indirect Dark Matter
detection signals varies over the allowed parameter space.

In the case at hand, even for $C=0$ particles can only annihilate with
anti--particles. For given total Dark Matter density and given
annihilation cross section the annihilation rate will therefore be two
times smaller than for self--conjugate (Majorana) Dark Matter. This is
essentially compensated by the fact that we need approximately two
times larger annihilation cross sections for $C=0$ to achieve a given
total Dark Matter density, compared to the case of Majorana dark
matter; see the discussion of Fig.~\ref{fig:e}. Fig.~\ref{fig:d} can
therefore also be used to compare with the more frequently studied
case of Majorana Dark Matter.

The left frame of Fig.~\ref{fig:d} shows the normalized product
$\sigma \cdot Y_\chi \cdot Y_{\bar\chi}$ as function of $m_\chi \cdot
C$, which again makes the result almost independent of $m_\chi$. We
also note that, as in the left frame of Fig.~\ref{fig:f}, the result
is almost identical for $a=0, \, b\neq 0$ and for $a\neq 0, \ b=0$;
this is again due to the fact that the annihilation cross section is
fixed, for given $C$, by the desired total Dark Matter relic density.

Recall that an increase of $C$ beyond zero corresponds to an increase
of the annihilation cross section, albeit initially a very small one
(see Fig.~\ref{fig:e}). This increases the first factor in the product
shown here. On the other hand, we just saw in Fig.~\ref{fig:f} that
this decreases the third factor, $Y_{\bar\chi}$. Fig.~\ref{fig:d}
shows that this second effect is always the more important one,
i.e. introducing an asymmetry in the Dark Matter sector can only {\em
  reduce} the signals from Dark Matter annihilation in galactic
haloes. However, this reduction is at first very mild. For constant
decoupling temperature and constant cross section the product $Y_\chi
\cdot Y_{\bar\chi}$ scales like $\left[Y_\chi(C=0)\right]^2 - C^2/4$,
which has vanishing slope at $C=0$. The increase of cross section
required by the increase of $C$ to keep the total Dark Matter density
fixed does not change this result significantly: the positive
contribution to the slope from the first factor in the product is
partly canceled by the resulting decrease of the decoupling
temperature, which reduces $Y_\chi(C=0)$. Recall also that initially
the required cross section depends only very weakly on $C$.

As a result of this small negative slope at small $C$, the signals
from WIMP annihilation in the halo still have about 20\% of the
strength in the symmetric case when the asymmetry $C$ reaches 95\% of
its maximal value. However, beyond that point the strength of these
signals plummet quickly, due to the very rapid decrease of the
$\bar\chi$ relic density which we already saw earlier.

The right frame of Fig.~\ref{fig:d} shows the normalized product of
annihilation cross section, $\chi$ density and $\bar\chi$ density as
function of $a + b/6$. We now observe an almost constant, negative
logarithmic slope as the cross section is increased from its minimal
value, which corresponds to $C=0$. Already for small $C$
the essentially vanishing slope in the left frame of Fig.~\ref{fig:d}
(product vs. $C$) combines with the very large positive slope of
Fig.~\ref{fig:e} (required $C$ vs. cross section) to give a finite
negative slope. 

In particular, we find that the indirect signals from WIMP
annihilation in galactic haloes are suppressed by about a factor
$10^5$ for a cross section about $8.2\ (9.0)$ times the minimal
allowed cross section, for $b=0 \ (a=0)$. This opens new possibilities
for model building. A suppression of the annihilation rate by a factor
${\cal O}(10^5)$ below the rate expected for self--conjugate thermal
dark matter annihilating from an $S-$wave initial state is required if
an observed excess of positrons near the galactic center \cite{posi}
is to be explained \cite{boehm} through the annihilation of WIMPs with
masses near an MeV.\footnote{Annihilation into $e^+e^-$ pairs is only
  possible if $m_\chi > m_e$. On the other hand, astrophysical
  observations \cite{bounds} imply $m_\chi \lsim 10$ MeV in this
  case.} In existing models \cite{boehm} this is engineered by
suppressing annihilation from the $S-$wave; the velocity dependence of
the cross section then suppresses today's annihilation rate relative
to that in the early universe. Here we see that the same suppression
can be arranged for asymmetric Dark Matter, if the annihilation cross
section is about one order of magnitude larger than that required for
symmetric DM.

\section{Summary and Conclusions}

In this paper we investigated the abundance of Dark Matter for the
asymmetric WIMP scenario, where Dark Matter particles and
anti--particles are distinct. This opens the possibility of generating
an asymmetry between WIMPs and their anti--particles. Here we assume
that this happened well before the epoch of thermal decoupling of the
WIMPs. We do not specify any dynamical mechanism for generating this
asymmetry; rather, we treat it as a free parameter.

In our work we have assumed that the Dark Matter has reached thermal
equilibrium before it decoupled. This allows a fairly generic
treatment of the problem. Asymmetric Dark Matter is often motivated by
explaining the origin of the baryon or lepton asymmetry. The
baryogenesis or leptogenesis process is usually assumed to happen much
earlier than WIMP decoupling. We note, however, that there could be
cases where the WIMP density did not reach full thermal
equilibrium. In that case, the abundance evolution would be
model--dependent. Moreover, we assume that WIMPs can only annihilate
with their anti--particles; self--annihilation of WIMPs would add
additional terms to the Boltzmann equations.

The Dark Matter decoupled from the rest of the cosmic fluid, and its
(co--moving) abundance froze, when the annihilation rate dropped below
the Hubble expansion rate. In the asymmetric Dark Matter case, as there
are more particles than anti--particles (for the reverse case, our
results would be the same except that ``particle'' and
``anti--particle'' are exchanged), the anti--particles annihilated
away more efficiently, with large numbers of particles left behind
without partner to annihilate. As a result, the final abundance is
determined not only by the cross section as in the symmetric Dark
Matter case, but also by the asymmetry.  We derived approximate
analytical expressions for the abundance and decoupling temperature as
functions of cross section and asymmetry, generalizing the well--known
treatment of the (chemical) decoupling of symmetric DM.

Numerically we find that the final DM density is already largely
determined by the initial asymmetry if the annihilation cross section
is just two times larger than the value required for symmetric thermal
DM. We also saw that introducing an asymmetry always reduces today's
indirect DM detection signals from WIMP annihilation in galactic
haloes: even though introducing an asymmetry requires larger
annihilation cross sections to attain the desired total DM relic
density, the resulting enhancement of the annihilation rate is
over--compensated by the strong suppression of the final anti--WIMP
density.  This suppression of today's WIMP annihilation rate is
desired in models where annihilation of MeV Dark Matter particles into
$e^+e^-$ pairs explains a possible excess of positrons found near the
center of our galaxy.

We find that increasing the annihilation cross section by one order of
magnitude suppresses today's annihilation rate by about six orders of
magnitude. Today's anti--WIMP density therefore remains significant
only for a rather narrow range of annihilation cross sections above
the value required for symmetric dark matter.

\subsubsection*{Note Added}
As we were preparing this manuscript, ref.~\cite{GSV} appeared, in
which the problem of asymmetric WIMP Dark Matter decoupling is also
investigated. While the general ideas are similar, the main focus of
the paper and the details of the treatment are different. In
particular, ref.~\cite{GSV} presents analytical expressions only for
the difference and the ratio of the (co--moving) $\chi$ and $\bar\chi$
densities. Numerically our results are in general agreement with theirs.

\section*{Acknowledgments}

We thank John Barrow for discussions. The work of H.I. is
supported by the National Natural Science Foundation of China
(11047009) and by the doctor fund BS100108 of Xinjiang
university. X.C. is supported by the Ministry of Science and
Technology National Basic Science Program (Project 973) under grant
No. 2007CB815401, and by the NSFC under grant No. 11073024. MD is
supported by the DFG TR33 `The Dark Universe.' He thanks the KIAS
School of Physics in Seoul as well as the Particle Theory Group at the
University of Hawaii at Manoa for hospitality.

\end{document}